\documentclass{article}
\usepackage[left=2cm, right=2cm, top=2cm]{geometry}
\usepackage{amssymb,amsmath,verbatim,mathtools,needspace,enumitem,etoolbox,graphicx,physics,microtype,afterpage,xspace,tabularx,lmodern,multirow}
\usepackage{authblk}
\usepackage{gensymb}
\usepackage[utf8]{inputenc}
\usepackage[usenames,dvipsnames]{xcolor}
\definecolor{linkcolor}{rgb}{0.0,0.3,0.5}
\definecolor{romared}{RGB}{142,0,28}
\usepackage[unicode, colorlinks=true, linkcolor=linkcolor, citecolor=blue, filecolor=linkcolor, urlcolor=linkcolor, linktocpage, breaklinks]{hyperref}
\usepackage[all]{hypcap}
\usepackage[T1]{fontenc}
\usepackage{subcaption}

\usepackage[russian, english]{babel}

\newcommand{\oxford}{Astrophysics, University of Oxford, Denys Wilkinson Building, Keble Road, Oxford OX1 3RH, United Kingdom}
\newcommand{\qmul}{Geometry, Analysis and Gravitation, School of Mathematical Sciences, Queen Mary University of London,
Mile End Road, London E1 4NS, United Kingdom}
\newcommand{\jhu}{Department of Physics and Astronomy, Johns Hopkins University, Baltimore, MD 21218, USA}
\newcommand{\potsdam}{Max Planck Institute for Gravitational Physics (Albert Einstein Institute), Am M\"uhlenberg 1, Potsdam-Golm, 14476, Germany}
\newcommand{\camb}{Department of Applied Mathematics and Theoretical Physics, 
Centre for Mathematical Sciences, University of Cambridge, Wilberforce Road, 
Cambridge CB3 0WA, United Kingdom}

\title{\textbf{GRDzhadzha: A code for evolving relativistic matter on analytic metric backgrounds}}
\author[1]{Josu C. Aurrekoetxea}
\author[1]{Jamie Bamber}
\author[2]{Sam E. Brady}
\author[2]{Katy Clough}
\author[3]{Thomas Helfer}
\author[1]{James Marsden}
\author[4]{Miren Radia}
\author[5]{Dina Traykova}
\author[3]{Zipeng Wang}
\date{}

\affil[1]{\oxford}
\affil[2]{\qmul}
\affil[3]{\jhu}
\affil[4]{\camb}
\affil[5]{\potsdam}

\begin{document}

\maketitle

The following brief overview has been prepared as part of the submission of the code to the Journal of Open Source Software. The code itself can be found at \url{https://github.com/GRChombo/GRDzhadzha}
\footnote{Dzhadzha (\foreignlanguage{russian}{джаджа}) is a word meaning gadget in Bulgarian.}.

\section{Summary}
Strong gravity environments such as those around black holes provide us with unique opportunities to study questions in fundamental physics, such as the existence and properties of dark matter and dark energy. Characterising the behaviour of new fields and other types of matter in highly relativistic environments generally necessitates numerical simulations unless one imposes significant symmetries.
Therefore we need to turn to numerical methods to study the dynamics and evolution of the complex systems of black holes and other compact objects in different environments, using numerical relativity (NR).
These methods allow us to split the four-dimensional Einstein equations into three-dimensional spatial hypersurfaces and a time-like direction.
Then if a solution is known at the initial spatial hypersurface, it can be numerically evolved in time, where an analytic solution no longer exists.
Whilst the tools of NR provide the most complete (i.e., approximation free) method for evolving matter in such environments, in many cases of interest, the density of the matter components is negligible in comparison to the curvature scales of the background spacetime metric, in which case it is a reasonable approximation to neglect their backreaction on it and treat the metric as fixed (assuming the background itself is stationary or otherwise has an analytic form).

In such cases, one does not need to evolve all the metric degrees of freedom as in NR, but only the additional matter ones. 
It is possible to do this using any NR code in a trivial way by setting the evolution of the metric variables to zero, but this is clearly rather inefficient. This code, GRDzhadzha, directly evolves the matter variables on an analytically specified background.
This significantly speeds up the computation time and reduces the resources needed (both in terms of CPU hours and storage) to perform a given simulation. 
The code is based on the publicly available NR code GRChombo \cite{Clough:2015sqa,Andrade:2021rbd}, which itself uses the open source Chombo framework \cite{Adams:2015kgr} for solving PDEs. In the following sections we discuss the key features and applications of the code, and give an indication of the efficiencies that can be achieved compared to a standard NR code.

\section{Key features}
GRDzhadzha inherits many of the features of GRChombo and Chombo, but avoids the complications introduced when evolving the metric. The key features are:
\begin{itemize}
    \item Background metrics: The currently available backgrounds in the code are a static Kerr black hole in horizon penetrating Kerr-Schild coordinates and a boosted black hole in isotropic Schwarzschild coordinates. These backgrounds can easily be adapted to other coordinate systems for different problems. The code is templated over the background so it can easily be changed without major code modification.
    \item Matter evolution: The code calculates the evolution for the matter variables on the metric background using an ADM decomposition in space and time - currently we have implemented a real and a complex scalar field as examples of matter types. Again the code is templated over the matter class so that the matter types can be exchanged with minimal modification.
    \item Accuracy: The metric values and their derivatives are calculated exactly at each point, whereas the matter fields are evolved with a 4th order Runge-Kutta time integration and their derivatives calculated with the same finite difference stencils used in GRChombo (4th and 6th order are currently available).
    \item Boundary Conditions: GRDzhadzha inherits all the available boundary conditions in GRChombo, namely, extrapolating, Sommerfeld (radiative), reflective and periodic. 
    \item Initial Conditions: The current examples provide initial data for real and complex scalar field matter. Since backreaction is ignored, there are no constraint equations to satisfy in the case of a scalar field, and the initial data can be freely specified.
    \item Diagnostics:  GRDzhadzha has routines for verifying the conservation of matter energy densities, angular and linear momentum densities, and their fluxes, as discussed in \cite{Clough:2021qlv,Croft:2022gks}.
    \item C++ class structure: Following the structure of GRChombo, GRDzhadzha is also written in C++ and uses object oriented programming (OOP) and templating.
    \item Parallelism: GRChombo uses hybrid OpenMP/MPI parallelism with explicit vectorisation of the evolution equations via intrinsics, and is AVX-512 compliant.
    \item Adaptive Mesh Refinement: The code inherits the flexible AMR grid structure of Chombo, which provides Berger-Oliger style \cite{Berger:1984zza} AMR with block-structured Berger-Rigoutsos grid generation \cite{Berger:1991}. Depending on the problem, the user may specify the refinement to be triggered by the matter or the background spacetime \cite{Radia:2021smk}. One nice feature is that one does not need to resolve the horizon of the black hole unless matter is present at that location, so for an incoming wave a lot of storage and processing time can be saved by only resolving the wave, and not the spacetime background.
\end{itemize}

\section{Statement of Need}
As mentioned in the introduction, any numerical relativity code like GRChombo can undertake these simulations. Examples of these include the Einstein Toolkit (\url{http://einsteintoolkit.org/}), with its related Cactus (\url{http://cactuscode.org}) \cite{Loffler:2011ay,Schnetter:2003rb}, and Kranc (\url{http://kranccode.org}) \cite{Husa:2004ip} infrastructure used by LEAN \cite{Sperhake:2006cy,Zilhao:2010sr} and Canuda (\url{https://bitbucket.org/canuda}) \cite{Witek:2018dmd}. Other notable but non-public codes include BAM \cite{Bruegmann:2006ulg,Marronetti:2007ya}, AMSS-NCKU \cite{Galaviz:2010mx}, PAMR/AMRD and HAD \cite{East:2011aa,Neilsen:2007ua}. Codes such as SPeC \cite{Pfeiffer:2002wt} and bamps \cite{Hilditch:2015aba} implement the generalised harmonic formulation of the Einstein equations using a pseudospectral method, and discontinuous Galerkin methods are used in SpECTRE (\url{https://spectre-code.org}) \cite{deppe_nils_2021_4734670,Kidder:2016hev,Cao:2018vhw}. NRPy (\url{http://astro.phys.wvu.edu/bhathome}) \cite{Ruchlin:2017com} is a code aimed for use on non-HPC systems, which generate C code from Python, and uses adapted coordinate systems to minimise computational costs. CosmoGRaPH (\url{https://cwru-pat.github.io/cosmograph}) \cite{Mertens:2015ttp} and GRAMSES \cite{Barrera-Hinojosa:2019mzo} are among several NR codes targeted at cosmological applications (see \cite{Adamek:2020jmr} for a comparison) and which also employ particle methods. Simflowny (\url{https://bitbucket.org/iac3/simflowny/wiki/Home}) \cite{Palenzuela:2018sly}, like CosmoGRaPH, is based on the SAMRAI infrastructure, and has targeted fluid and MHD applications. GRAthena++ \cite{Daszuta:2021ecf} makes use of oct-tree AMR to maximise scaling.

Whilst there exist many NR codes (both public and private), which can in principle be used to perform simulations of fundamental fields on a fixed BH background, most do not have the efficiency advantages of GRDzhadzha\footnote{As far as we are aware, only NRPy \cite{Ruchlin:2017com} and Canuda \cite{Witek:2018dmd} offer the same functionality. Some private codes also have such capabilities (see e.g.~\cite{Traykova:2017zrn}, based on \cite{Braden:2014cra}). Other codes may have similar features that are not explicitly separated out, so this makes it difficult to identify them.}.
In particular, the fact that the ADM variables and their derivatives are not evolved or stored on the grid saves both a lot of simulation run time, as well as output file storage space.
To get a rough idea of the improvement in storage and CPU hours one can achieve, we performed a short test simulation using GRDzhadzha and compared it simulation performed using the full NR capabilities of GRChombo.
We find that on average GRDzhadzha is 15-20 times faster than GRChombo and requires about 3 times less file storage.
An additional advantage of this code versus using a full NR code, for problems with negligible backreaction, is that here the metric variables are calculated analytically at every point on the grid, which significantly decreases the margin for numerical error, and means that resolution can be focussed on the matter location, and not the spacetime curvature.

It is important to note that whilst backreaction is neglected in the metric calculation, this does not mean that the backreaction effects cannot be calculated. Fixed background simulations provide a first order (in the density) estimate of the gravitational effects caused by the matter, taking into account their relativistic behaviour. This is discussed further in \cite{Clough:2021qlv} and some examples using the approach are \cite{Bamber:2020bpu, Traykova:2021dua, Traykova:2023qyv}. 

Since the interface and structure of the code is very close to the GRChombo numerical relativity code, it is possible for the results of these fixed background simulations to be used as initial data in full numerical relativity simulations (and vice versa), as was done in \cite{Bamber:2022pbs}. Therefore if the backreaction is found to be significant due to some growth mechanism, the simulation can be continued in full NR.

\section{Research projects using GRDzhadzha}

\begin{figure}[t!]
\centering
\begin{subfigure}{0.48\linewidth}
  \includegraphics[width=\linewidth]{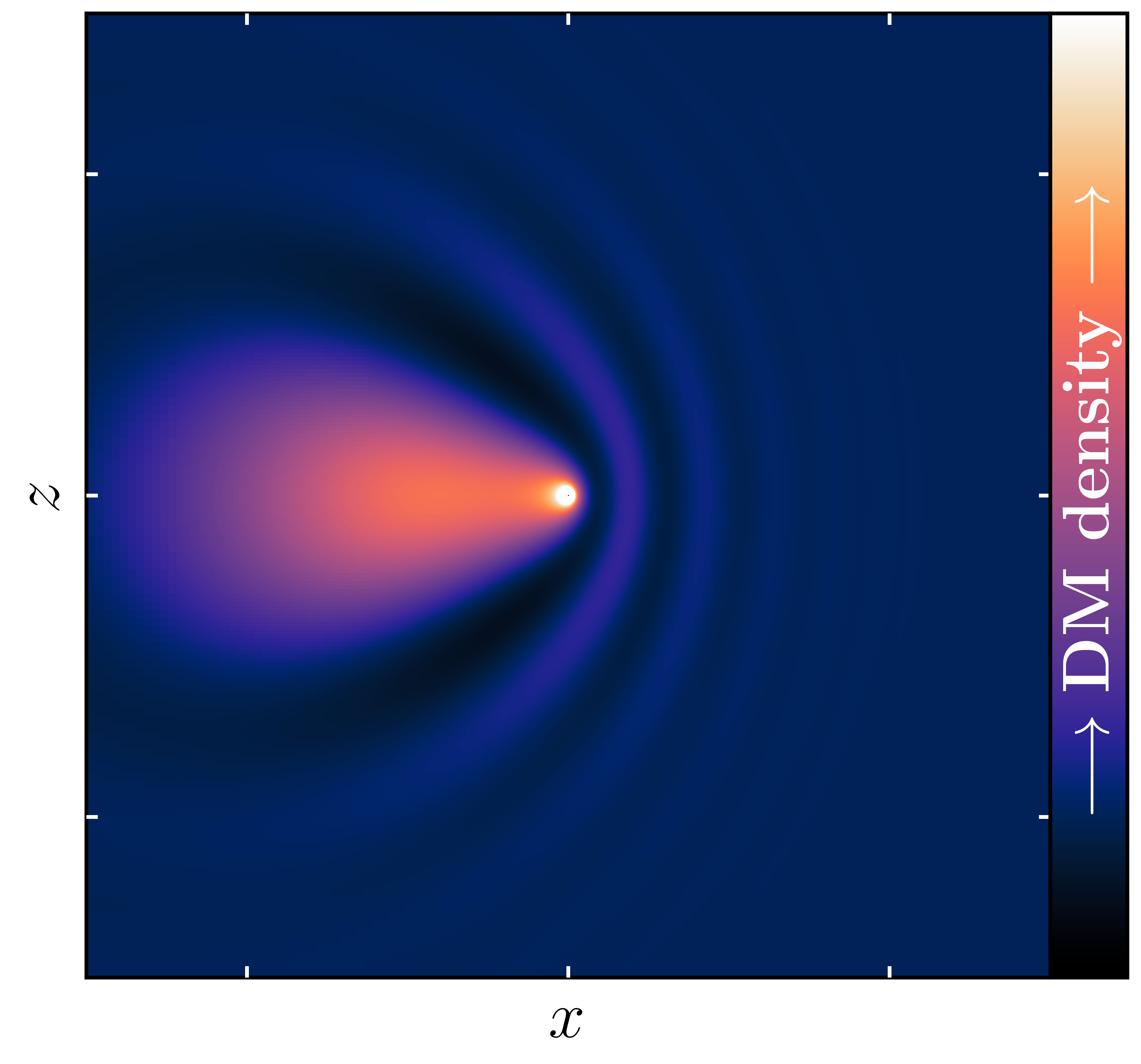}
  \label{fig:sub1}
\end{subfigure}%
\hspace{0.03\linewidth}
\begin{subfigure}{0.48\linewidth}
  \includegraphics[width=\linewidth]{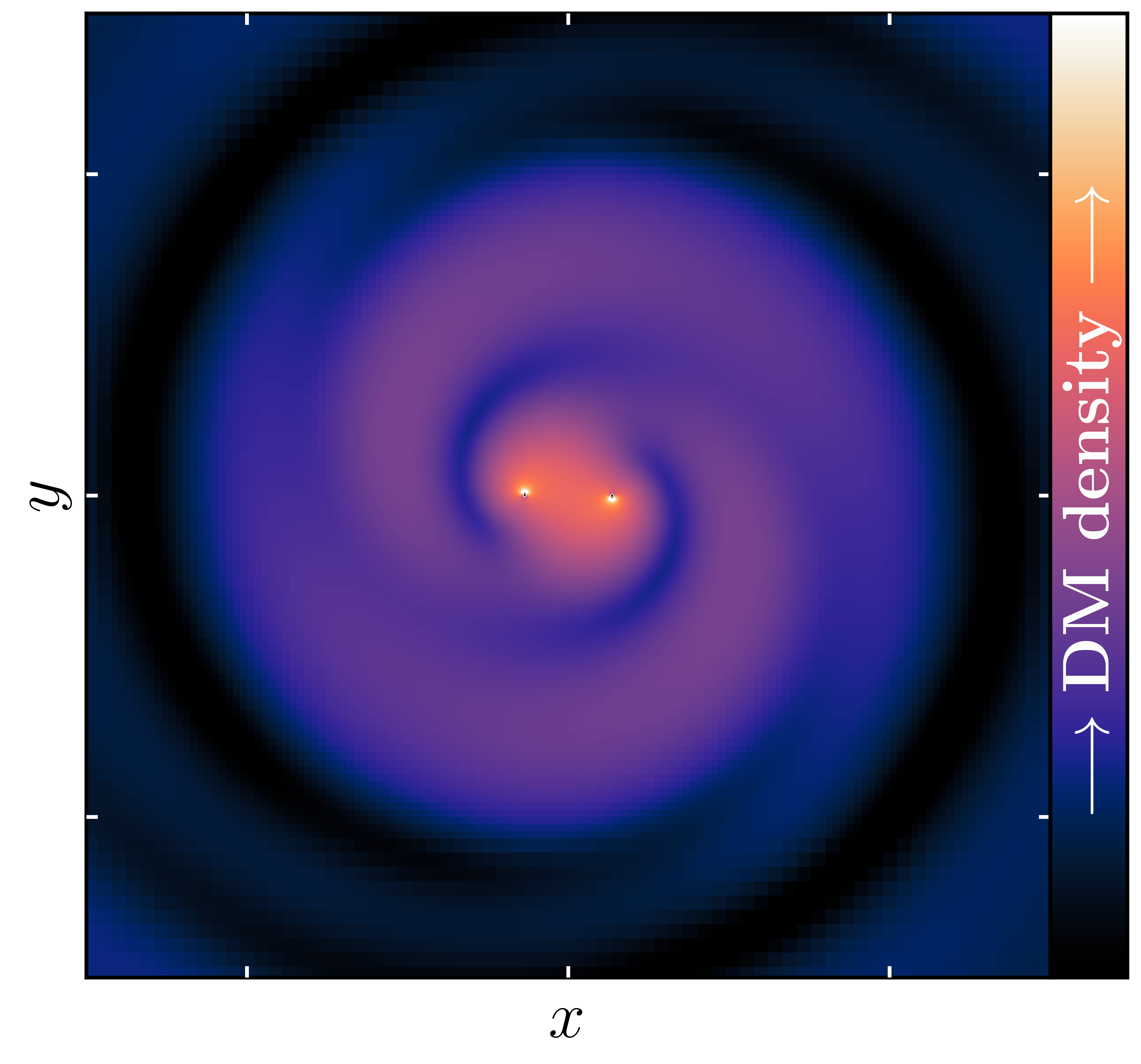}
  \label{fig:sub2}
\end{subfigure}
\vspace{-25pt}
\caption{Some examples of the physics studied with GRDzhadzha. The left image is taken from \cite{Traykova:2021dua} in which the dynamical friction of light dark matter was studied in the relativistic regime. The right image is from a study of the scalar clouds around black hole binaries in \cite{Bamber:2022pbs}, in which the initial conditions were generated with a modified version of GRDzhadzha.}
\end{figure}

So far the code has been used to study a range of fundamental physics problems:
\begin{itemize}
    \item Studying the interference patterns in neutrino flavour oscillations around a static black hole \cite{Alexandre:2018crg}.
    \item Growth of scalar hair around a Schwarzschild \cite{Clough:2019jpm} and a Kerr \cite{Bamber:2020bpu} black hole.
    \item Determining the relativistic drag forces on a Schwarzschild black hole moving through a cloud of scalar field dark matter \cite{Traykova:2021dua, Traykova:2023qyv}.  
    \item Studying the dynamical friction effects on a Kerr black hole (Magnus effect) \cite{Wang:inprep}.
    \item Superradiance with self-interacting vector field \cite{Clough:2022ygm} and with spatially varying mass \cite{Wang:2022hra}.
    \item BH mergers in wave dark matter environments \cite{Bamber:2022pbs}.
\end{itemize}

\section*{Acknowledgements}

\noindent We thank the GRChombo collaboration (\href{www.grchombo.org}{www.grchombo.org}) for their support and code development work. JB and JM acknowledge funding from a UK Science and Technology Facilities Council (STFC) studentship.
JCA acknowledges funding from the Beecroft Trust and The Queen’s College via an extraordinary Junior Research Fellowship (eJRF).
A few of the projects using this code have received funding from the European Research Council (ERC) under the European Union’s Horizon 2020 research and innovation programme (Grant Agreement No 693024).
KC acknowledges funding from the UKRI Ernest Rutherford Fellowship (grant number ST/V003240/1). 
SEB is supported by a QMUL Principal studentship.
GRDzhadzha users have benefited from the provision of HPC resources from: 

\begin{itemize}
    \item DiRAC (Distributed Research utilising Advanced Computing) resources under the projects ACSP218, ACSP191, ACTP183, ACTP186 and ACTP316. DiRAC is part of the National e-Infrastructure. Systems used include: 
    \begin{itemize}
        \item Cambridge Service for Data Driven Discovery (CSD3), part of which is operated by the University of Cambridge Research Computing on behalf of the STFC DiRAC HPC Facility (www.dirac.ac.uk). The DiRAC component of CSD3 was funded by BEIS capital funding via STFC capital grants ST/P002307/1 and ST/R002452/1 and STFC operations grant ST/R00689X/1

        \item DiRAC Data Intensive service at Leicester, operated by the University of Leicester IT Services, which forms part of the STFC DiRAC HPC Facility (www.dirac.ac.uk). The equipment was funded by BEIS capital funding via STFC capital grants ST/K000373/1 and ST/R002363/1 and STFC DiRAC Operations grant ST/R001014/1
        
        \item DiRAC at Durham facility managed by the Institute for Computational Cosmology on behalf of the STFC DiRAC HPC Facility (www.dirac.ac.uk). The equipment was funded by BEIS capital funding via STFC capital grants ST/P002293/1 and ST/R002371/1, Durham University and STFC operations grant ST/R000832/1

        \item DiRAC Complexity system, operated by the University of Leicester IT Services, which forms part of the STFC DiRAC HPC Facility (www.dirac.ac.uk). This equipment is funded by BIS National E-Infrastructure capital grant ST/K000373/1 and STFC DiRAC Operations grant ST/K0003259/1
    \end{itemize}
    
  \item Sakura, Cobra and Raven clusters at the Max Planck Computing and Data Facility (MPCDF) in Garching, Germany

  \item PRACE (Partnership for Advanced Computing in Europe) resources under grant numbers 2018194669, 2020225359. Systems used include:
    \begin{itemize}
        \item SuperMUCNG, Leibniz Supercomputing Center (LRZ), Germany
        \item JUWELS, Juelich Supercomputing Centre (JSC), Germany
    \end{itemize}
    
    \item the Argonne Leadership Computing Facility, including the Joint Laboratory for System Evaluation (JLSE), which is a U.S. Department of Energy (DOE) Office of Science User Facility supported under Contract DE-AC02-06CH11357

  \item the Texas Advanced Computing Center (TACC) at the University of Austin HPC and visualization resources URL: 
  (\href{http://www.tacc.utexas.edu}{http://www.tacc.utexas.edu}), and the San Diego Supercomputing Center (SDSC) URL: (\href{https://www.sdsc.edu}{https://www.sdsc.edu}), under project PHY-20043 and XSEDE Grant No. NSF-PHY-090003

  \item the Glamdring cluster, Astrophysics, Oxford, UK
\end{itemize}

\bibliographystyle{vancouver}
\bibliography{main}

\end{document}